\documentstyle[sprocl]{article}
\def\lref#1#2{{\bibitem{#1}#2}}

\def\np#1#2#3{Nucl. Phys. {\bf B#1} (#2), #3}
\def\pl#1#2#3{Phys. Lett. {\bf #1B} (#2), #3}

\def\pr#1#2#3{Phys. Rev. {\bf #1} (#2), #3}
\def\aph#1#2#3{Ann. Phys. {\bf #1} (#2), #3}

\def\cmp#1#2#3{Comm. Math. Phys. {\bf #1} (#2), #3}

\def\jhep#1#2#3{JHEP {\bf#1}(#2), #3}

\def\atmp#1#2#3{Adv.~Theor.~Math.~Phys.{\bf #1} (#2), #3}
\def\ap#1#2#3{Ann.~Phys. {\bf #1} (#2), #3}
\def\IB{\relax\hbox{$\inbar\kern-.3em{\rm B}$}}
\def\IC{{\bf C}}
\def\ID{\relax\hbox{$\inbar\kern-.3em{\rm D}$}}
\def\IE{\relax\hbox{$\inbar\kern-.3em{\rm E}$}}
\def\IF{\relax\hbox{$\inbar\kern-.3em{\rm F}$}}
\def\IG{\relax\hbox{$\inbar\kern-.3em{\rm G}$}}
\def\IGa{\relax\hbox{${\rm I}\kern-.18em\Gamma$}}
\def\IH{\relax{\rm I\kern-.18em H}}
\def\IK{\relax{\rm I\kern-.18em K}}
\def\IL{\relax{\rm I\kern-.18em L}}
\def\IP{\relax{\rm I\kern-.18em P}}
\def\IR{\relax{\rm I\kern-.18em R}}
\def\IZ{\relax\ifmmode\mathchoice{
\hbox{\cmss Z\kern-.4em Z}}{\hbox{\cmss Z\kern-.4em Z}}
                 {\lower.9pt\hbox{\cmsss Z\kern-.4em Z}}
                {\lower1.2pt\hbox{\cmsss Z\kern-.4em Z}}
                          \else{\cmss Z\kern-.4em Z}\fi}
\def\II{\relax{\rm I\kern-.18em I}}

\def\ndt{{\noindent}}

\def\CA{{\cal A}}

\def\CD{{\cal D}}
\def\CE{{\cal E}}
\def\CF{{\cal F}}
\def\CG{{\cal G}}
\def\CH{{\cal H}}
\def\CI{{\cal I}}

\def\CM{{\cal M}}
\def\CN{{\cal N}}
\def\CO{{\cal O}}

\def\p{{\partial}}

\def\nb{\bar{n}}

\def\zb{\bar{z}}

\def\half{{\scriptscriptstyle{1 \over 2}}}

\def\Tr{{\rm Tr}}
\def\Id{{\rm Id}}

\def\inbar{\,\vrule height1.5ex width.4pt depth0pt}
\font\cmss=cmss10 \font\cmsss=cmss10 at 7pt

\def\a{{\alpha}}
\def\ap{{{\a}^{\prime}}}
\def\b{{\beta}}
\def\d{{\delta}}

\def\e{{\epsilon}}
\def\z{{\zeta}}
\def\ve{{\varepsilon}}

\def\m{{\mu}}
\def\n{{\nu}}

\def\k{{\kappa}}
\def\l{{\lambda}}
\def\s{{\sigma}}
\def\t{{\theta}}

\def\nct{{noncommutative \,}}
\def\be{\begin{equation}}
\def\ee{\end{equation}}
\def\bea{\begin{eqnarray}}
\def\eea{\end{eqnarray}}
\def\crt{\nonumber \\}
\def\eqn#1#2{{\be #2 \label{#1} \ee}}

\def\ctn#1{{{\cite{#1}}}}
\def\rfn#1{{({\ref{#1}})}}
\begin{document}
\title{NONCOMMUTATIVE INSTANTONS REVISITED}
\author{Nikita A. Nekrasov }
\address{Institute for Theoretical and Experimental
Physics, 117259 Moscow, Russia\\ Joseph Henry Laboratories,
Princeton University, Princeton, New Jersey 08544\\ IHES, Le
Bois-Marie, 35 route de Chartres, Bures-sur-Yvette, F-91440,
France\\} \maketitle \abstracts{We find a new gauge in which U(1)
noncommutative instantons are explicitly non-singular on
noncommutative ${\bf R}^4$. We also present a pedagogical introduction into
noncommutative gauge theories.}
\section{Introduction}

Recently there has been a revival of interest in noncommutative
gauge theories\ctn{snyder}, \ctn{connes}. They are interesting as examples of field theories
which have as their planar limit large $N$ gauge theories
\ctn{planar},\ctn{filk}; certain supersymmetric versions of \nct
gauge theories  arise as ${\ap} \to 0$ limit of theories on
Dp-branes in the presence of background $B$-field
\ctn{douglashull},\ctn{witsei}; the related theories arise in
Matrix compactifications with $C$-field turned on \ctn{cds}; finally, noncommutativity is in some sense an intrinsic feature of the open string field
theory \ctn{wtnc}, \ctn{k}, \ctn{volker}. 
A lot of progress has been recently achieved in the 
analysis of the classical solutions of the \nct gauge theory. 
The first explicit solutions and their moduli where analysed in \ctn{neksch}
where instantons in the four dimensional \nct gauge theory (with self-dual 
noncommutativity) were constructed. These instantons play an important role
in the construction of the discrete light cone quantization of the
M-theory fivebrane \ctn{abkss}, \ctn{abs}, and they also gave a hope of 
giving an interpretation in 
the physical gauge theory language of 
the torsion free sheaves  which appear in various 
interpretations of D-brane states \ctn{avatars}, \ctn{harveymoore}, 
in particular
those responsible for the enthropy of black holes realized via D5-D1 systems
\ctn{vafastrominger}, 
and also entering the S-duality invariant partition functions of ${\CN}=4$
super-Yang-Mills theory \ctn{vafawitten}.
In addition to the instantons (which are particles in 4+1 dimensional theory),
which represent the D0-D4 system, 
the monopole-like solutions were found\ctn{grossnek} in U(1) gauge theory in 3+1 dimensions.
The latter turn out to have a string attached to them. Both the string
and the monopole at its end are the \nct filed theory realization of the
D3-D1 system, where D1 string ends on the D3 brane and bends at some specific
angle towards the brane. 
One can also find the solutions describing the string itself 
\ctn{alexios}, \ctn{grossneki}, both the BPS and in the 
non-BPS states; also the dimensionally reduced solutions in 2+1 dimensions
\ctn{grossneki}, \ctn{amgs}, describing the D0-D2 systems; finite length
strings, corresponding to $U(2)$ monopoles \ctn{grossnekii}.

This paper is devoted to 
the clarification of the issue of nonsingularity of 
the \nct U(1) instantons. We shall show that one can find a 
gauge in which the 
solutions are explicitly nonsingular, 
and well-defined over all of \nct ${\IR}^4$. 
Compared to \ctn{neksch} we also relax the assumption on the noncommutativity. We shall only demand that the Poisson tensor ${\t}^{ij}$ has
non-negative Pfaffian: ${\rm Pf}({\t}) \neq 0$ 
(and of course, that the space is noncommutative, i.e. at least one of the 
eigen-values 
of ${\t}^{ij}G_{jk}{\t}^{kl}$ 
must be non-vanishing, $G$ being the Euclidean metric on the space). 

The paper is organized as follows. The section $2$ contains a pedagogical introduction into \nct gauge theories. The section $3$ constructs instantons in \nct
gauge theory on ${\bf R}^4$ for any group $U(N)$. The section $4$
presents explicit formulae for the $U(1)$ gauge group.

{\bf Note added.} As this paper was ready for publication two related 
papers appeared.
The paper \ctn{amgs} also discusses codimension four solitons in 
\nct gauge theory, using operators $S, S^{\dagger}$ (which we introduce later in section $4$) 
. These, however, are non-BPS solutions (and the role of $S$ and $S^{\dagger}$
is reversed), and do not obey instanton
equations. 
The paper \ctn{fkii} overlaps with us in that it also 
uses of the operators $S, S^{\dagger}$ 
for constructing instanton gauge fields. 
Also, some of the discussion of the relation of the
torsion free sheaves on ${\IC}^2$ to the \nct instantons is similar.

\section{Noncommutative Geometry and Noncommutative Field Theory}

\subsection{A brief mathematical introduction}

It has been widely appreciated by the mathematicians (starting
with the seminal works of  Gelfand, Grothendieck, and von Neumann)
that the geometrical properties of a space $X$ are encoded in the
properties  of the commutative algebra $C(X)$ of the continuous
functions $f: X \to {\IC}$ with the ordinary rules of point-wise
addition and multiplication: $(f + g)(x) = f(x) + g(x), f\cdot g
(x) = f(x) g(x)$.

More precisely, $C(X)$ knows only about the topology of $X$, but
one can refine the definitions and look at the algebra
$C^{\infty}(X)$ of the smooth functions or even at the DeRham
complex ${\Omega}^{\cdot}(X)$ to decipher the geometry of $X$.

The algebra $A = C(X)$ is clearly associative, commutative and has
a unit (${\bf 1} (x) = 1$). It also has an involution, which maps
a function to its complex conjugate: $f^{\dagger} (x) =
\overline{f(x)}$.

The points $x$ of $X$ can be viewed in two ways: as maximal ideals
of $A$: $f \in I_{x} \Leftrightarrow f(x) = 0$; or as the
irreducible (and therefore one-dimensional for $A$ is commutative)
representations of $A$: $R_x (f) = f(x)$, $R_x \approx {\IC}$.

The vector bundles over $X$ give rise to projective modules over
$A$. Given a bundle $E$ let us consider the space ${\CE} =
{\Gamma} (E)$ of its sections. If $f \in A$ and ${\s} \in {\CE}$
then clearly $f {\s} \in {\CE}$. This makes ${\CE}$ a
representation of $A$, i.e. a module. Not every module over $A$
arises in this way. The vector bundles over topological spaces
have the following remarkable property, which is the content of
Serre-Swan theorem: for every vector bundle $E$ there exists
another bundle $E^{\prime}$ such that the direct sum $E \oplus
E^{\prime}$ is a trivial bundle $X \times {\IC}^{N}$ for
sufficiently large $N$. When translated to the language of modules
this property reads as follows: for the module ${\CE}$ over $A$
there exists another module ${\CE}^{\prime}$ such that ${\CE}
\oplus {\CE}^{\prime} = F_{N} = A^{\oplus N}$. We have denoted by
$F_{N} = A \otimes_{\IC} {\IC}^{N}$ the free module over $A$ of
rank $N$. Unless otherwise stated the symbol ${\otimes}$ below
will be used for tensor products over ${\IC}$. The modules with
this property are called {\it projective}. The reason for them to
be called in such a way is that ${\CE}$ is an image of the free
module $F_{N}$ under the projection which is identity on ${\CE}$
and zero on ${\CE}^{\prime}$. In other words, for each projective
module ${\CE}$ there exists $N$ and an operator $P \in {\rm Hom}
(F_{N}, F_{N})$, such that $P^2 = P$, and ${\CE} = P \cdot F_{N}$.

Noncommutative geometry relaxes the condition that $A$ must be
commutative, and tries to develop a geometrical intuition about
the noncommutative associative algebras with anti-holomorphic
involution $^{\dagger}$ (${\IC}^*$-algebras).

In particular, the notion of vector bundle over $X$ is replaced by
the notion of the projective module over $A$. Now, when $A$ is
noncommutative, there are two kinds of modules: left and right
ones. The left $A$-module is the vector space $M_{l}$ with the
operation of left multiplication by the elements of the algebra
$A$: for $m \in M_l$ and  $a \in A$ there must be an element $a m
\in M_l$, such that for $a_1, a_2$: $a_1 (a_2 m) = (a_1 a_2) m$.
The definition of the right $A$-module $M_r$ is similar: for $m
\in M_r$ and  $a \in A$ there must be an element $m a\in M_r$,
such that for $a_1, a_2$: $(m a_1) a_2  = m (a_1 a_2)$. The free
module $F_N = A \oplus \ldots_{\scriptscriptstyle N \, \rm
times}\oplus A = A \otimes {\IC}^{N}$ is both left and right one.
The projective $A$-modules are defined just as in the commutative
case, except that for the left projective $A$-module ${\CE}$ the
module ${\CE}^{\prime}$, such that ${\CE} \oplus {\CE}^{\prime} =
F_N$, also must be left, and similarly for the right modules.

The manifolds can be mapped one to another by means of smooth
maps: $g : X_1 \to X_2$. The algebras of smooth functions  are
mapped in the opposite way: $g^* : C^{\infty}(X_2) \to
C^{\infty}(X_1)$, $g^* (f) (x_1) = f ( g(x_1))$. The induced map
of the algebras is the algebra homomorphism: $$g^* (f_1 f_2) =
g^*(f_1) g^*(f_2), \, g^* (f_1 + f_2) = g^* ( f_1) + g^* (f_2)$$

Naturally, the smooth maps between two manifolds are
replaced by the homomorphisms of the corresponding algebras. In particular,
the maps of the manifold to itself form the associative algebra
$Hom (A, A)$. The diffeomorphisms would correspond to the invertible
homomorphisms, i.e. automorphisms $Aut (A)$. Among those there are internal,
generated by the invertible elements of the algebra:
$$
a \mapsto g^{-1} a g
$$
The infinitesimal diffeomorphisms of the ordinary
manifolds are generated by the vector fields $V^i {\p}_i$, which
 differentiate
functions, $$f \mapsto f + {\ve} V^i {\p}_i f $$ In the
noncommutative setup the vector field is replaced by the
derivation of the algebra $V \in Der (A)$: $$ a \mapsto a + {\ve}
V(a), \quad V( a) \in A$$ and the condition that $V(a)$ generates
an infinitesimal homomorphism reads as: $$V (a b) = V (a) b + a
V(b)$$ which is just the definition of the derivation. Among
various derivations there are internal ones, generated by the
elements of the algebra itself: $$V_{c}(a) = [ a, c] : = a c - c
a, \quad c \in A$$ These infinitesimal diffeomorphisms are absent
in the commutative setup, but they have close relatives in the
case of Poisson manifold $X$.

\subsection{Flat noncommutative space}

The basic example of the noncommutative algebra which will be
studied here is the enveloping algebra of the Heisenberg algebra.
Consider the Euclidean space ${\bf R}^{d}$ with coordinates $x^i$,
$i=1, \ldots, d$. Suppose a constant antisymmetric matrix
${\t}^{ij}$ is fixed. It defines a Poisson bi-vector field
${\t}^{ij} {\p}_i {\p}_j$ and therefore the noncommutative
associative product on ${\bf R}^{d}$. The coordinate functions
$x^i$ on the deformed noncommutative manifold will obey the
following commutation relations: \eqn{cmrl}{[x^i, x^j] = i
{\t}^{ij}\ , } We shall call the algebra ${\CA}_{\t}$ (over
${\IC}$) generated by the $x^i$ satisfying \rfn{cmrl}, together
with  convergence conditions on the allowed expressions of the
$x^i$ -- the noncommutative space-time. The algebra ${\CA}_{\t}$
has an involution $a \mapsto a^{\dagger}$ which acts as a complex
conjugation on the central elements $\left( {\l} \cdot {\bf 1}
\right)^{\dagger} = {\bar\l} \cdot {\bf 1}, \, \l \in {\IC}$ and
preserves $x^i$: $(x^i)^{\dagger} = x^i$. The elements of
${\CA}_{\t}$ can be identified with ordinary complex-valued
functions on ${\bf R}^d$, with the product of two functions $f$
and $g$ given by the Moyal formula (or star product): \eqn{myl}{f
\star g \, (x)= {\exp} \left[ {i \over 2} {\t}^{ij} {{\p}\over{\p
x_{1}^{i}}} {{\p}\over{\p x_{2}^{j}}} \right] f (x_{1}) g (x_{2})
\vert_{x_{1} = x_{2} = x}\ .}

\subsubsection{Fock space formalism.}

By an orthogonal change of coordinates we can map the Poisson
tensor ${\t}_{ij}$ onto its canonical form: $$ x^i \mapsto z_a,
{\zb}_{a}, \quad a = 1, \ldots, r\ ; \quad y_{b}, \quad b = 1,
\ldots, d- 2r ,$$ so that: \eqn{ncm}{[y_a,y_b]= [y_b, z_a] = [y_b,
{\zb_a}] = 0, \quad [z_{a}, {\zb}_{b}] = -2{\t}_{a}{\d}_{ab} ,
{\quad} {\t}_{a}
> 0} $$ ds^2 =dx^2_i+ dy_b^2 = dz_a d{\zb}_{a} + dy_b^2 . $$
{}Since $z (\zb)$ satisfy (up to a constant) the commutation
relations of creation (annihilation) operators we can identify
functions $f(x,y)$ with  the functions of the $y_a$ valued in the
space of operators acting in the Fock space ${\CH}_r$ of $r$
creation and annihilation operators: \eqn{fock}{{\CH}_r =
\bigoplus_{\vec n}\, {\IC} \, \vert n_1, \ldots, n_r \rangle} $$
c_{a} = {1\over \sqrt{2{\t}_a}} {\zb}_a , \quad c^{\dagger}_a =
{1\over \sqrt{2\t_a}} z_{a}, \,  [c_{a}, c^{\dagger}_{b}] =
{\d}_{ab} $$ $$ c_a \vert {\vec n} \rangle = \sqrt{n_{a}} \vert
{\vec n}-1_{a} \rangle, \quad c_a^{\dagger} \vert {\vec n} \rangle
= \sqrt{n_{a} + 1} \vert {\vec n} + 1_a \rangle $$ Let ${\hat
n}_{a} = c^{\dagger}_{a} c_{a}$ be the $a$'th number operator.

The Hilbert space ${\CH}_r$ is the example of left projective
module over the algebra ${\CA}_{\t}$. Indeed, consider the element
$P_{0} = \vert \vec 0 \rangle \langle \vec 0 \vert \sim {\exp} -
\sum_{a} {{z_a {\zb}_a}\over{{\t}_a}}$. It obeys $P_0^2 = P_0$,
i.e. it is a projector. Consider the rank one free module $F_{1} =
{\CA}_{\t}$ and let us consider its left sub-module, spanned by
the elements of the form: $f \star P_0$. As a module it is clearly
isomorphic to ${\CH}_r$, isomorphism being: ${\vert \vec n
\rangle} \mapsto {\vert \vec n \rangle \langle
 \vec 0 \vert}$. It is projective module, the complementary
module being ${\CA}_{\t} ( 1 - P_0) \subset {\CA}_{\t}$.

{}The procedure that maps ordinary commutative functions onto
operators in the Fock space acted on by $z_a, {\zb}_a$ is called
Weyl ordering and is defined by: \eqn{wlor}{f(x) \mapsto {\hat
f(z_a, {\bar z}_a)} =  \int f(x) \, {{{\rm d}^{2r} x \,\, {\rm
d}^{2r} p }\over{(2{\pi})^{2r}}} \,
  \,  e^{ i \left( {\bar p}_a z_a + p_{a} {\zb}_a -  p \cdot x \right)}.} It is easy to see
that \eqn{product}{  {\rm if } \quad f \mapsto \hat f, \quad g
\mapsto \hat g \quad {\rm then }\quad f\star g \mapsto \hat f \hat
g .}

\subsubsection{Symmetries of the flat noncommutative space}

The algebra \rfn{cmrl} has an obvious symmetry: $x^i \mapsto x^i +
{\ve}^i$, for ${\ve}^i \in {\bf R}$. For invertible Poisson
structure ${\t}$ it is an example of the internal automorphism of
the algebra: \eqn{auto}{a \mapsto e^{ i{\t}_{ij} {\ve}^{i} x^{j}}
a e^{-i{\t}_{ij}{\ve}^{i}x^j}} In addition, there are rotational
symmetries which we shall not need.

\subsection{Gauge theory on  noncommutative space}

{}In an  ordinary gauge theory with gauge group $G$ the gauge
fields are connections in some principal $G$-bundle. The matter
fields are the sections of the vector bundles with the structure
group $G$. Sections of the noncommutative vector bundles are
elements of the projective modules over the algebra ${\CA}_{\t}$.

{}In the ordinary gauge theory the gauge field arises through the
operation of covariant differentiation of the sections of a vector
bundle.  In the \nct setup the situation is similar. Suppose $M$
is a projective module over ${\CA}$. The connection ${\nabla}$ is
the operator $$ {\nabla} : {\bf R}^d \times M \to M, \quad
{\nabla}_{\ve} (m) \in M, \qquad {\ve} \in {\bf R}^d, \, m \in M \
,  $$ where ${\bf R}^d$ denotes the commutative vector space, the
Lie algebra of the automorphism group generated by \rfn{auto}. The
connection is required to obey the Leibnitz rule:
\eqn{leibn}{{\nabla}_{\ve} ( a m_{\bf l} ) = {\ve}^i ( {\p}_i a )
m_{\bf l} + a {\nabla}_{\ve} m_{\bf l}} \eqn{reibn}{{\nabla}_{\ve}
( m_{\bf r} a ) = m_{\bf r} {\ve}^i ( {\p}_i a ) + (
{\nabla}_{\ve} m_{\bf r} ) a \ . } Here, \rfn{leibn} is the
condition for left modules, and \rfn{reibn} is the condition for
the right modules. {}As usual, one defines the curvature $F_{ij} =
[{\nabla}_i, {\nabla}_j]$ - the operator ${\Lambda}^2 {\bf R}^d
\times M \to M$ which commutes with the multiplication by $a \in
{\CA}_{\t}$. In other words, $F_{ij} \in {\rm End}_{\CA}(M)$. In
ordinary gauge theories the gauge fields come with gauge
transformations. In the \nct case the gauge transformations, just
like the gauge fields, depend on the module they act in. For the
module $M$ the group of gauge transformations ${\CG}_{M}$ consists
of the invertible endomorphisms of $M$ which commute with the action of
${\CA}$ on $M$: $${\CG}_{M} = {\rm GL}_{\CA} (M)$$

All the discussion above can be specified to the case where the
module has a Hermitian inner product, with values in ${\CA}$.

\subsubsection{Fock module and connections there.}

Recall that the algebra ${\CA}_{\t}$ for $d = 2r$ and
non-degenerate ${\t}$ has an important irreducible representation,
the left module ${\CH}_{r}$. Let us now ask, what kind of
connections does the module ${\CH}_{r}$ have?

By definition, we are looking for a collection of operators
${\nabla}_{i} : {\CH}_r \to {\CH}_r$, $i=1, \ldots, 2r$, such
that: $$ [ {\nabla}_i, a ] = {\p}_i a $$ for any $a \in {\CA}$.
Using the fact that ${\p}_i a = i{\t}_{ij} [ x^j, a] $ and the
irreducibility of ${\CH}_r$ we conclude that:
\eqn{cnfc}{{\nabla}_i = i {\t}_{ij} x^j + {\k}_i, \qquad {\k}_i
\in {\IC}} If we insist on unitarity of ${\nabla}$, then $i {\k}_i
\in {\bf R}$. Thus, the space of all gauge fields suitable for
acting in the Fock module is rather thin, and is isomorphic to the
vector space ${\bf R}^{d}$ (which is canonically dual to the Lie
algebra of the automorphisms of ${\CA}_{\t}$). The gauge group for
the Fock module, again due to its irreducibility is simply the
group $U(1)$, which multiplies all the vectors in ${\CG}_r$ by a
single phase. In particular, it preserves ${\k}_i$'s, so they are
gauge invariant. {}It remains to find out what is the curvature of
the gauge field given by \rfn{cnfc}. The straightforward
computation of the commutators gives: \eqn{fccrvt}{F_{ij} = i
{\t}_{ij}} i.e. all connections in the Fock module have the
constant curvature.

\subsubsection{Free modules and connections there.}

If the right (left) module $M$ is free, i.e. it is a sum of
several copies of the algebra ${\CA}_{\t}$ itself, then the
connection ${\nabla}_i$ can be written as $$ {\nabla}_i = {\p}_i +
A_i $$ where $A_i$ is the operator of the left (right)
multiplication by the matrix with ${\CA}_{\t}$-valued entries:
\eqn{conct}{{\nabla}_i m_{\bf l} = {\p}_i m_{\bf l} + m_{\bf l}
A_i, \, {\nabla}_i m_{\bf r} = {\p}_i m_{\bf r} + A_i m_{\bf r}}In
the same operator sense the curvature obeys the standard identity:
$$ F_{ij} = {\p}_i A_j - {\p}_j A_i + A_i A_j - A_j A_i \ . $$
{}Given a module $M$ over some algebra $A$ one can multiply it by
a free module $A^{\oplus N}$ to make it a module over an algebra
${\rm Mat}_{N \times N}(A)$ of matrices with elements from $A$. In
the non-abelian gauge theory over $A$ we are interested in
projective  modules over ${\rm Mat}_{N \times N}(A)$. If the
algebra $A$ (or perhaps its subalgebra) has a trace, ${\Tr}$, then
the algebra ${\rm Mat}_{N \times N} (A)$ has a trace given by the
composition of a usual matrix trace and ${\Tr}$.

{}It is a peculiar property of the noncommutative algebras that
the algebras $A$ and ${\rm Mat}_{N \times N}(A)$ have much in
common. These algebras are called Morita equivalent and under some
additional conditions the gauge theories over $A$ and over ${\rm
Mat}_{N\times N}(A)$ are also equivalent. This phenomenon is
responsible for the similarity between the "abelian
noncommutative" and "non-abelian commutative" theories.

{}If we represent ${\p}_i$ as $i {\t}_{ij} [ x^j, \cdot] $ then
the expression for the covariant derivative becomes:
\eqn{back}{{\nabla}_i m_{\bf l} = i {\t}_{ij} x^j m_{\bf r} +
m_{\bf r} D_i, \quad {\nabla}_i m_{\bf r} =  - m_{\bf r}i
{\t}_{ij}x^j +  D_{i}^{\dagger} m_{\bf r}} where \eqn{backi}{D_i =
- i{\t}_{ij} x^j + A_i}

\vfill\eject\section{Instantons in noncommutative gauge theories}

{}We would like to  study the non-perturbative objects in
noncommutative gauge theory.

{}Specifically we shall be interested in four dimensional
instantons. They either appear as instantons themselves in the
Euclidean version of the four dimensional theory (theory on
Euclidean D3-brane), as solitonic particles in the theory on
D4-brane, i.e. in 4+1 dimensions, or as instanton strings in the
theory on D5-brane (and are related to little strings
\ctn{lttlstr}). 
They also show up as 
``freckles'' in the gauge theory/sigma model correspondence \ctn{freck}.

{}The theory depends on the dimensionfull parameters ${\t}_{\a}$
which enter the commutation relation between the coordinates of
the space: $[x , x] \sim i {\t}$.

We treat  only the bosonic fields,
but these could  be a part of a supersymmetric multiplet, with
${\CN}=2$ supersymmetry or higher. Such field theories arise on
the world volume of D3-branes in the presence of a background
constant $B$-field along the D3-brane.

A D3-brane can be surrounded by other branes as well. For example,
in the Euclidean setup, a D-instanton could approach the D3-brane.
In fact, unless the D-instanton is dissolved inside
   the brane, the combined system breaks supersymmetry \ctn{witsei}. The
D3-D(-1) system can be rather simply described in terms of  a
noncommutative $U(1)$ gauge theory - the latter has instanton-like
solutions \ctn{neksch}. It is the purpose of this note to explore
these solutions in greater detail.

{}More generally, one can have a stack of $k$ Euclidean D3-branes
with $N$ D(~-1)-branes inside. This configuration will be
described by charge $N$ instantons in $U(k)$ gauge theory.

Let us work in four Euclidean space-time dimensions, ${\m} =
1,2,3,4$. As we said above, we shall look at the purely bosonic
Yang-Mills theory on the space-time ${\CA}_{\t}$ with the
coordinates functions $x^{\m}$ obeying the Heisenberg commutation
relations: \eqn{he}{[ x^{\m}, x^{\n} ] = i {\t}^{\m\n}} We assume
that the metric on the space-time is Euclidean: \eqn{mtr}{ds^2 =
\sum_{\m} (dx^{\m})^2} The action describing our gauge theory is
given by: \eqn{lgrn}{S = - {1\over{4g_{\rm YM}^2}}  {\Tr} F \wedge
\star F \,  + {{\t}\over{8\pi^2}}  {\Tr} F \wedge F} where $g_{\rm
YM}^2$ is the Yang-Mills coupling constant, and \eqn{crvt}{F =
F_{\m\n} dx^{\m} \wedge dx^{\n}, \quad F_{\m\n} = {\nabla}_{\m}
{\nabla}_{\n} - {\nabla}_{\n} {\nabla}_{\m}} The covariant
derivatives ${\nabla}_{\m}$ act in some module ${\CE}$ over the
algebra ${\CA}_{\t}$ of functions on the noncommutative ${\bf
R}^4$.

\subsection{Instantons}

\ndt {}The equations of motion following from \rfn{lgrn} are:
\eqn{eom}{{\nabla}_{\m} F_{\m\n} = 0} In general these equations
are as hard to solve as the equations of motion of the ordinary
non-abelian Yang-Mills theory. However, just like in the
commutative case, there are special solutions, which are simpler
to analyze and which play a crucial role in the analysis of the
quantum theory. These are the so-called (anti-)instantons. The
(anti-)instantons solve the first order equation:
\eqn{asd}{F_{\m\n} = {\pm} {1\over{2}} {\ve}_{\m\n\k\l} F_{\k\l}}
These equations are easier to solve. The solutions are classified
by the instanton charge: \eqn{inch}{N = - {1\over{8\pi^2}} {\Tr} F
\wedge F}

\subsection{ADHM construction}

 In the commutative case all
solutions to \rfn{asd} with the finite action \rfn{lgrn} are
obtained via the so-called Atiyah-Drinfeld-Hitchin-Manin (ADHM)
construction.  If we are concerned with the instantons in the
$U(k)$ gauge group, then the ADHM data consists of
\begin{enumerate}
\item the pair of the two complex vector spaces $V$ and $W$
of dimensions $N$ and $k$ respectively;

\item the operators: $B_{1}, B_{2} \in {\rm Hom} ( V, V)$, and
$I \in {\rm Hom} (W, V)$, $J \in {\rm Hom}(V, W)$;

\item the dual gauge group $G_{N} = U(N)$, which acts on the data above as
follows: \eqn{dgga}{B_{\a} \mapsto g^{-1} B_{\a} g, \,\, I \mapsto
g^{-1} I, \,\, J \mapsto J g}

\item Hyperk\"ahler quotient\ctn{rkh} with respect to the group \rfn{dgga}. It
means that one takes the set $X_{k,N} = {\m}_{r}^{-1}(0) \cap
{\m}_{c}^{-1}(0)$ of the common zeroes of the three moment maps:
\bea & {\m}_{r} = [B_{1} , B_{1}^{\dagger}] + [ B_{2},
B_{2}^{\dagger}] + II^{\dagger} - J^{\dagger}J \crt & {\m}_{c} =
[B_{1}, B_{2}] + IJ \\ & {\bar\m}_{c} = [ B_{2}^{\dagger},
B_{1}^{\dagger}] + J^{\dagger} I^{\dagger} \nonumber \label{mmnt}
\eea\\ and quotients it by the action of  $G_{N}$.
\end{enumerate}
The claim of ADHM is that the points in the space ${\CM}_{k,N} =
X_{k,N}^{\circ} /G_{N}$ parameterize the solutions to \rfn{asd}
(for ${\t} =0$ ) up to the gauge transformations. Here
$X^{\circ}_{k,N} \subset X_{k,N}$ is the open dense subset of
$X_{k,N}$ which consists of the solutions to ${\vec \m} =0$ such
that their stabilizer in $G_{N}$ is trivial. The explicit formula
for the gauge field $A_{\m}$ is also known. Define the Dirac-like
operator: \eqn{drc}{{\CD}^{+}  = \pmatrix{ -B_2 + z_2 & B_1 - z_1
& I \cr B_1^{\dagger} - {\zb}_1 & B_2^{\dagger} - {\zb}_2 &
J^{\dagger} \cr}: V \otimes {\IC}^2 \oplus W \to V \otimes
{\IC}^2} Here $z_1, z_2$ denote the complex coordinates on the
space-time: $$ z_1 = x_1 + i x_2,  \quad z_2 = x_3 + i x_4, \quad
{\zb}_1 = x_1 - ix_2, \quad {\zb}_2 = x_3 - ix_4 $$ The kernel of
the operator \rfn{drc} is the $x$-dependent vector space
${\CE}_{x} \subset  V \otimes {\IC}^2 {\oplus} W$. For generic
$x$, ${\CE}_{x}$ is isomorphic to $W$. Let us denote by ${\Psi} =
{\Psi}(x)$ this isomorphism. In plain words, ${\Psi}$ is the
fundamental solution to the equation: \eqn{fnde}{{\CD}^{+} {\Psi}
= 0, \qquad {\Psi}: W \to V \otimes {\IC}^2 {\oplus} W} If the
rank of ${\Psi}$ is $x$-independent (this property holds for
generic points in $\CM$), then one can normalize:
\eqn{nrml}{{\Psi}^{\dagger} {\Psi} = {\Id}_{k}}which fixes $\Psi$
uniquely up to an $x$-dependent $U(k)$ transformation ${\Psi}(x)
\mapsto {\Psi}(x) g (x)$, $g(x) \in U(k)$. Given $\Psi$ the
anti-self-dual gauge field is constructed simply as follows:
\eqn{asdg}{{\nabla}_{\m} = {\p}_{\m} + A_{\m}, \qquad A_{\m} =
{\Psi}^{\dagger}(x) {{\p}\over{{\p} x^{\m}}} {\Psi} (x)} The space
of $(B_{0}, B_{1}, I, J)$ for which $\Psi (x)$ has maximal rank
for all $x$ is an open dense subset ${\CM}_{N,k} =
X^{\circ}_{N,k}/G_{N}$ in ${\CM}$. The rest of the points in
$X_{N,k}/G_{N}$ describes the so-called {\it point-like}
instantons. Namely, ${\Psi} (x)$ has maximal rank for all $x$ but
some finite set $\{ x_1, \ldots, x_l \}$, $l \leq k$. The
\rfn{nrml} holds for $x \neq x_i, \, i = 1, \ldots, l$, where the
left hand side of \rfn{nrml} simply vanishes.

{}The noncommutative deformation of the gauge theory leads to the
noncommutative deformation of the ADHM construction. It turns out
to be very simple yet surprising. The same data $V, W, B, I, J,
\ldots$ is used. The deformed ADHM equations are simply
\eqn{dfadhm}{{\m}_r = {\z}_r, \, {\m}_c = {\z}_c}where we have
introduced the following notations. The Poisson tensor ${\t}^{ij}$
entering the commutation relation $[x^i, x^j] = i{\t}^{ij}$ can be
decomposed into the self-dual and anti-self-dual parts
${\t}^{\pm}$. If we look at the commutation relations of the
complex coordinates $z_1, z_2, {\zb}_1, {\zb}_2$ then the
self-dual part of ${\t}$ enters the following commutators:
\eqn{cmtrs}{[z_1, z_2] = - {\z}_c \qquad [z_1, {\zb}_1] + [z_2,
{\zb}_2]  =  - {\z}_r} It turns out that as long as $\vert \z
\vert = {\z}_r^2 + {\z}_c {\bar\z}_c
>0$ one needs not to distinguish between ${\widetilde X}_{N,k}$
and ${\widetilde X}_{N,k}^{\circ}$, in other words the stabilizer
of any point in ${\widetilde X}_{N,k} = {\m}_{r}^{-1}(-{\z}_{r})
\cap {\m}_{c}^{-1}(-{\z}_{c})$ is trivial. Then the resolved
moduli space is ${\widetilde\CM}_{N,k} = {\widetilde
X}_{N,k}/G_{N}$.

By making an orthogonal rotation on the coordinates $x^{\m}$ we
can map the algebra ${\CA}_{\t}$ onto the sum of two copies of the
Heisenberg algebra. These two algebras can have different values
of ``Planck constants''. Their sum is the norm of the self-dual
part of ${\t}$, i.e. $\vert {\z} \vert$, and their difference is
the norm of anti-self-dual part of ${\t}$: \eqn{tcha}{[z_1,
{\zb}_1] = - {\z}_1, \, \quad [z_2, {\zb}_2] = -{\z}_2}where
${\z}_1 + {\z}_2 = \vert {\t}^{+} \vert$, ${\z}_1 - {\z}_2 = \vert
{\t}^{-}\vert$. By the additional reflection of the coordinates,
if necessary, one can make both ${\z}_1$ and ${\z}_2$ positive
(however, one should be careful, since if the odd number of
reflections was made, then the orientation of the space was
changed and the notions of the instantons and anti-instantons are
exchanged as well).

{}The next step in the ADHM construction was the definition of the
isomorphism ${\Psi}$ between the fixed vector space $W$ and the
fiber ${\CE}_{x}$ of the gauge bundle, defined as the kernel of
the operator ${\CD}^{+}$. In the noncommutative setup one can also
define the operator ${\CD}^{+}$ by the same formula \rfn{drc}. It
is a map between two free modules over ${\CA}_{\t}$:
\eqn{frm}{{\CD}_{x}^{+}: \left( V \otimes {\IC}^2 \oplus W \right)
\bigotimes {\CA}_{\t} \to \left( V \otimes {\IC}^2 \right)
\bigotimes {\CA}_{\t}} which commutes with the right action of
${\CA}_{\t}$ on the free modules. Clearly, $$ {\CE} = {\rm Ker}
{\CD}^{+}$$ is a right module over ${\CA}_{\t}$, for if ${\CD}^{+}
s = 0$, then ${\CD}^{+} (s \cdot a) = 0$, for any $a \in
{\CA}_{\t}$.

${\CE}$ is also a projective module, for the following reason.
Consider the operator ${\CD}^{+}{\CD}$. It is a map from the free
module $V \otimes {\IC}^2 \bigotimes {\CA}_{\t}$ to itself. Thanks
to \rfn{dfadhm} this map actually equals to ${\Delta} \otimes
{\Id}_{{\IC}^2}$ where ${\Delta}$ is the following map from the
free module $V \otimes {\CA}_{\t}$ to itself: \eqn{dlta}{\Delta =
(B_1 - z_1) (B_1^{\dagger} - {\zb}_1) + (B_2 - z_2 )(B_2^{\dagger}
- {\zb}_2) + II^{\dagger}} We claim that ${\Delta}$ has no kernel,
i.e. no solutions to the equation ${\Delta} \, v = 0$, $v \in V
\otimes {\CA}_{\t}$. Recall the Fock space representation ${\CH}$
of the algebra ${\CA}_{\t}$. The coordinates $z_{\a}, {\zb}_{\a}$,
obeying \rfn{tcha}, with ${\z}_1, {\z}_2 > 0$,  are represented as
follows: \eqn{rp}{z_1 = \sqrt{{\z}_1} \, c_1^{\dagger}, \, {\zb}_1
= \sqrt{{\z}_1} \, c_1 , \quad z_2 = \sqrt{{\z}_2} \,
c_2^{\dagger}, \, {\zb}_2 = \sqrt{{\z}_2} \, c_2} where $c_{1,2}$
are the annihilation operators and $c_{1,2}^{\dagger}$ are the
creation operators for the two-oscillators Fock space $${\CH} =
\bigoplus_{\scriptscriptstyle n_1, n_2 \geq 0} {\IC} \, \vert n_1,
n_2 \rangle$$ Let us assume the opposite, namely that there exists
a vector $v \in V \otimes {\CA}_{\t}$ such that ${\Delta} v = 0$.
Let us act by this vector on an arbitrary state $\vert n_1, n_2
\rangle$ in ${\CH}$. The result is the vector ${\n}_{\nb} \in V
\otimes {\CH}$ which must be annihilated by the operator
${\Delta}$, acting in $V \otimes {\CH}$ via \rfn{rp}. By taking
the Hermitian inner product of the equation ${\Delta} {\n}_{\nb}
=0$ with the conjugate vector ${\n}_{\nb}^{\dagger}$ we
immediately derive the following three equations: $$
(B_2^{\dagger} - {\zb}_2 ) {\n}_{\nb} \quad
 = \quad 0 $$ \eqn{triur}{(B_1^{\dagger} - {\zb}_1 ) {\n}_{\nb} \quad  =
\quad 0 } $$ I^{\dagger} {\n}_{\nb} \quad  = \quad 0 $$Using
\rfn{dfadhm} we can also represent ${\Delta}_x$ in the form:
\eqn{dltai}{{\Delta} = (B_1^{\dagger} - {\zb}_1) (B_1 - z_1) +
(B_2^{\dagger} - {\zb}_2) (B_2 - z_2) + J^{\dagger}J \ .} From
this representation another triple of equations follows: $$(B_2 -
z_2 ) {\n}_{\nb} \quad  = \quad 0 $$ \eqn{triura}{ (B_1 - z_1 )
{\n}_{\nb} \quad  = \quad 0} $$J {\n}_{\nb} \quad  = \quad 0 $$
Let us denote by $e_i$, $i=1, \ldots, N$ some orthonormal basis in
$V$. We can expand ${\n}_{\nb}$ in this basis as follows: $$
{\n}_{\nb} = \sum_{i=1}^{N} e_{i} \otimes v^i_{\nb}, \qquad
v^i_{\nb} \in {\CH}$$ The equations \rfn{triur},\rfn{triura}
imply: \eqn{repr}{(B_{\a})^i_j v^j_{\nb} = z_{\a} v^i_{\nb}, \quad
(B_{\a}^{\dagger})^{i}_j v^j_{\nb} = {\zb}_{\a} v^{i}_{\nb},
\qquad {\a} = 1,2} in other words the matrices $B_{\a},
B_{\a}^{\dagger}$ form a finite-dimensional representation of the
Heisenberg algebra which is impossible if either ${\z}_{1}$ or
${\z}_2 \neq 0$. Hence ${\n}_{\nb} =0$, for any ${\nb} = (n_1,
n_2)$ which implies that $v =0$.

Thus the Hermitian operator ${\Delta}$ is invertible. It allows to
prove the following theorem: each vector ${\psi}$ in the free
module $( V \otimes {\IC}^2 \oplus W) \otimes {\CA}_{\t}$ can be
decomposed as a sum of two orthogonal vectors: \eqn{dcmps}{ {\psi}
=
{\Psi}_{\psi} \oplus {\CD} {\chi}_{\psi}, \qquad {\CD}^{+}
{\Psi}_{\psi} = 0, \quad {\chi}_{\psi} \in (V \otimes {\IC}^2)
\otimes {\CA}_{\t}} where the orthogonality is understood in the
sense of the following ${\CA}_{\t}$-valued Hermitian product: $$
\langle \psi_1, \psi_2 \rangle = {\Tr}_{\scriptstyle V \otimes
{\IC}^2 \oplus W} \quad \left( {\psi}_1^{\dagger} {\psi}_2
\right)$$ The component ${\Psi}_{\psi}$ is annihilated by
${\CD}^{+}$, that is ${\Psi}_{\psi} \in {\CE}$. The image of
${\CD}$ is another right module over ${\CA}$ (being the image of
the free module $(V \otimes {\IC}^2 ) \otimes {\CA}_{\t}$): $$
{\CE}^{\prime}  = {\CD} (V \otimes {\IC}^2 \otimes {\CA}_{\t})$$
and their sum is a free module: $$ {\CE} \oplus {\CE}^{\prime} =
{\CF} := \left( V \otimes {\IC}^2 \oplus W \right) \otimes
{\CA}_{\t}$$ hence ${\CE}$ is projective. It remains to give the
expressions for ${\Psi}_{\psi}, {\chi}_{\psi}$:
\eqn{pro}{{\chi}_{\psi} = {1\over {\CD}^{+}{\CD}} {\CD}^{+}
{\psi}, \quad {\Psi}_{\psi} = \Pi {\psi}, \quad {\Pi} = \left( 1 -
{\CD}{1\over {\CD}^{+}{\CD}} {\CD}^{+} \right)}The noncommutative
instanton is a connection in the module ${\CE}$ which is obtained
simply by projecting the trivial connection on the free module
${\CF}$ down to ${\CE}$. To get the covariant derivative of a
section $s \in {\CE}$ we view this section as a section of
${\CF}$, differentiate it using the ordinary derivatives on
${\CA}_{\t}$ and project the result down to ${\CE}$ again:
\eqn{cvrnt}{{\nabla} s = {\Pi} \, {\rm d} s} The curvature is
defined through ${\nabla}^2$, as usual: \eqn{crvtr}{{\nabla}
{\nabla} s = F \cdot s = {\rm d} {\Pi} \wedge {\rm d} {\Pi} \cdot
s} where we used the following relations: \eqn{prjctn}{{\Pi}^2 =
{\Pi}, \quad {\Pi} s = s} Let us now show explicitly that the
curvature \rfn{crvtr} is anti-self-dual, i.e.
\eqn{asdcnd}{[{\nabla}_{\m}, {\nabla}_{\n} ] + {\half}
{\e}_{\m\n\l\rho} [ {\nabla}_{\l}, {\nabla}_{\rho}] = 0} First we
prove the following lemma: for any $s \in {\CE}$: \eqn{lemm}{{\rm
d} {\Pi} \wedge {\rm d} {\Pi} s = {\Pi} {\rm d} {\CD}
{1\over{{\CD}^{+} {\CD}}} {\rm d} {\CD}^{+} s} Indeed, $${\rm d}
{\Pi} \wedge {\rm d} {\Pi} = {\rm d} \left( {\CD}
{1\over{{\CD}^{+} {\CD}}} {\CD}^{+} \right)  \wedge {\rm d} \left(
{\CD} {1\over{{\CD}^{+} {\CD}}} {\CD}^{+} \right), $$ $$ {\rm d}
\left( {\CD} {1\over{{\CD}^{+} {\CD}}} {\CD}^{+} \right)
 = {\Pi} {\rm d} {\CD} {1\over{{\CD}^{+} {\CD}}} {\CD}^{+}
 +  {\CD} {1\over{{\CD}^{+} {\CD}}} {\rm d} {\CD}^{+} {\Pi},
$$ $$ {\CD}^{+} {\Pi} = 0 $$ hence $$
 {\rm d} \left( {\CD} {1\over{{\CD}^{+}
{\CD}}} {\CD}^{+} \right)  \wedge  {\rm d} \left( {\CD}
{1\over{{\CD}^{+} {\CD}}} {\CD}^{+} \right) =
\qquad\qquad\qquad$$ $$ \qquad\qquad\qquad\qquad = {\Pi} {\rm d}
{\CD} {1\over{{\CD}^{+}{\CD}}} {\rm d} {\CD}^{+} {\Pi} +
{\CD} {1\over{{\CD}^{+}{\CD}}} {\rm d} {\CD}^{+}\, {\Pi}\,
{\rm d} {\CD} {1\over{{\CD}^{+}{\CD}}} {\CD}^{+} $$ and
the second term vanishes when acting on $s \in {\CE}$, while the
first term gives exactly what the equation \rfn{lemm} states.

Now we can compute the curvature more or less explicitly:
\eqn{crva}{F\cdot  s = 2 {\Pi} \pmatrix{{1\over{\Delta}} f_3 &
{1\over{\Delta}} f_+ & 0 \cr {1\over{\Delta}} f_{-} & -
{1\over{\Delta}} f_3  & 0 \cr 0 & 0 & 0 } \cdot s} where $f_{3},
f_{+}, f_{-}$ are the basic anti-self-dual two-forms on ${\bf
R}^4$: \eqn{sdf}{f_3 = {\half} \left( {\rm d} z_1 \wedge {\rm d}
{\zb}_1 - {\rm d} z_2 \wedge {\rm d} {\zb}_2 \right),\, f_{+}  =
{\rm d} z_1 \wedge {\rm d} {\zb}_2, \, f_{-}= {\rm d} {\zb}_1
\wedge {\rm d} z_2}

Thus we have constructed the nonsingular anti-self-dual gauge
fields over ${\CA}_{\t}$. The interesting feature of the
construction is that it produces the non-trivial modules over the
algebra ${\CA}_{\t}$, which are projective for any point in the
moduli space. This feature is lacking in the ${\z} \to 0$, where
it is spoiled by the point-like instantons. This feature is also
lacking if the deformed ADHM equations are used for construction
of gauge bundles directly over a commutative space. In this case
it turns out that one can construct a torsion free sheaf over
${\IC}^2$, which sometimes can be identified with a holomorphic
bundle. However, generically this sheaf will not be locally free.
It can be made locally free by blowing up sufficiently many points
on ${\IC}^2$, thereby effectively changing the topology of the
space \ctn{branek}. The topology change is rather mysterious if we
recall that it is purely gauge theory we are dealing with.
However, in the supersymmetric case this gauge theory is an ${\ap}
\to 0$ limit of the theory on a stack of Euclidean D3-branes. One
could think that the topology changes reflect the changes of
topology of D3-branes embedded into flat ambient space. This is
indeed the case for monopole solutions, e.g.
\ctn{curtjuan},\ctn{hashimoto},\ctn{moriyama},\ctn{mateos}. It is
not completely unimaginable possibility, but so far it has not
been justified (besides from the fact that the DBI solutions
\ctn{witsei}, \ctn{terashima} are ill-defined without a blowup of
the space). What makes this unlikely is the fact that the
instanton backgrounds have no worldvolume scalars turned on. 

At any rate, the noncommutative instantons constructed above are
well-defined and nonsingular without any topology change.

Also note, that we have constructed instantons for arbitrary
noncommutativity tensor ${\t}_{\m\n}$, the only requirement being
the positivity of the Pfaffian ${\rm Pf}({\t}) \propto {\z}_1
{\z}_2 > 0$ (for ${\rm Pf}({\t}) < 0$ our formulae define
anti-instantons). For

\subsection{The identificator $\Psi$}

In the \nct case one can also try to construct the identifying map
${\Psi}$. It is to be thought as of the homomorphism of the
modules over ${\CA}$: $$ {\Psi}: W \otimes {\CA}_{\t} \to {\CE}$$
The normalization \rfn{nrml}, if obeyed, would imply the unitary
isomorphism between the free module $W \otimes {\CA}_{\t}$ and
${\CE}$. We can write: ${\Pi} = {\Psi}{\Psi}^{\dagger}$ and the
elements $s$ of the module ${\CE}$ can be cast in the form:
\eqn{sec}{s = {\Psi}\cdot {\s}, \qquad {\s} \in W \otimes
{\CA}_{\t}} Then the covariant derivative can be written as:
\eqn{cvidnt}{{\nabla} s = {\Pi} d ({\Psi} \cdot{\s}) = {\Psi}
{\Psi}^{\dagger} d \left({\Psi}{\s} \right) = {\Psi} \left( d {\s}
+ A {\s} \right)} where \eqn{ggfld}{A = {\Psi}^{\dagger} d {\Psi}}
just like in the commutative case. Introducing the {\it background
independent} operators $D_{\m} = i {\t}_{\m\n}x^{\n} + A_{\m}$, we
can write: \eqn{bcind}{D_{\m} = i {\Psi}^{\dagger}
{\t}_{\m\n}x^{\n} {\Psi}}

\section{Abelian instantons}

Let us describe the case of $U(1)$ instantons in detail.
In our notations above we have: $k=1$. It is known, from \ctn{nakajima},
that for ${\z}_r > 0, {\z}_c = 0$ the solutions to the deformed
ADHM equations have $J=0$. Let us denote by $V$ the complex
Hermitian vector space of dimensionality $N$, where $B_{\a}$, ${\a}= 1, 2$
act. Then $I$ is identified with a vector in $V$.
We can choose our units and coordinates in such a way that ${\z}_r = 2, {\z}_c = 0$.

\subsection{Torsion free sheaves on ${\bf C}^2$}

Let us recall at this point the algebraic-geometric interpretation
of the space $V$ and the triple $(B_{1}, B_{2}, I)$. The space
${\tilde X}_{N,1}$ parameterizes the rank one torsion free sheaves
on ${\IC}^2$. In the case of ${\bf C}^2$ these are identified with
the ideals ${\CI}$ in the algebra ${\IC}[z_1, z_2]$ of holomorphic
functions on ${\IC}^2$, such that $V = {\IC}[z_1, z_2]/{\CI}$ has
dimension $N$. An ideal of the algebra ${\CO} \approx {\IC}[z_1,
z_2]$ is a subspace ${\CI} \subset {\CO}$, which is invariant
under the multiplication by the elements of ${\CO}$, i.e. if $g
\in {\CI}$ then $f g \in {\CI}$ for any ${\CO}$.

An example of such an ideal is given by the space of functions of
the form: $$ f (z_1, z_2) = z_1^{N} g (z_1, z_2) + z_2 h(z_1, z_2)
$$The operators $B_{\a}$ are simply the operations of
multiplication of a function, representing an element of $V$ by
the coordinate function $z_{\a}$, and the vector $I$ is the image
in $V$ of the constant function $f = 1$. In the example above,
following\ctn{branek} we identify $V$ with ${\IC}[z_1]/z_1^{N}$,
the operator $B_2 = 0$, and in the basis $e_i = \sqrt{(i-1)!}
z_1^{N-i}$ the operator $B_1$ is represented by a Jordan-type
block: $B_1 e_i = \sqrt{2(i-1)} e_{i-1}$,  and $I = \sqrt{2N}
e_{N}$.

Conversely, given a triple $(B_1, B_2, I)$, such that the ADHM
equations are obeyed the ideal ${\CI}$ is reconstructed as
follows. The polynomial $f \in {\IC}[z_1, z_2]$ belongs to the
ideal, $f \in {\CI}$ if and only if $f(B_1, B_2) I = 0$. Then,
from the ADHM equations it follows that by acting on the vector
$I$ with polynomials in $B_1, B_2$ one generates the whole of $V$.
Hence ${\IC}[z_1, z_2]/{\CI} \approx V$ and has dimension $N$, as
required.

\subsection{Identificator $\Psi$ and projector $P$}
Let us now solve the equations for the identificator:
${\CD}^{\dagger}{\Psi}= 0, \, {\Psi}^{\dagger}{\Psi} = 1$. We
decompose: \eqn{dciden}{\Psi = \pmatrix{{\psi}_{+} \cr {\psi}_{-}
\cr {\xi}}} where ${\psi}_{\pm} \in V \otimes {\CA}_{\t}$, ${\xi}
\in {\CA}_{\t}$. The normalization \rfn{nrml} is now:
\eqn{anrml}{{\psi}^{\dagger}_{+}{\psi}_{+} +
{\psi}^{\dagger}_{-}{\psi}_{-} + {\xi}^{\dagger}{\xi} = 1}It is
convenient to work with rescaled matrices $B$:
 $B_{\a} = \sqrt{\z_{\a}} {\b}_{\a}$, ${\a}= 1,2$.
The equation ${\CD}^{\dagger}{\Psi}=0$ is solved by the
substitution: \eqn{ansatz}{{\psi}_{+}=  - \sqrt{{\z}_2} (
{\b}_2^{\dagger} - c_{2}) v, \quad {\psi}_{-} = \sqrt{{\z}_1} (
{\b}_{1}^{\dagger} - c_1 ) v} provided \eqn{master}{{\hat\Delta} v
+ I {\xi} = 0, \qquad {\hat\Delta} = \sum_{\a} {\z}_{\a}
({\b}_{\a} - c^{\dagger}_{\a})({\b}_{\a}^{\dagger} -
c_{\a})}Fredholm's alternative states that the solution ${\xi}$ of
\rfn{master} must have the property, that for any  ${\n} \in
{\CH}, {\chi} \in V$, such that \eqn{babush}{{\hat\Delta} ({\n}
\otimes {\chi}) = 0,} the equation \eqn{fred}{\left(
{\n}^{\dagger}\otimes {\chi}^{\dagger} \right) I {\xi} = 0}holds.
It is easy to describe the space of all ${\n} \otimes {\chi}$
obeying \rfn{babush}: it is spanned by the vectors:
\eqn{spnd}{e^{\sum {\b}_{\a}^{\dagger} c_{\a}^{\dagger}} \vert
0,0\rangle \otimes e_i, \quad i = 1, \ldots, N} where $e_i$ is any
basis in $V$. Let us introduce a Hermitian operator $G$ in $V$:
\eqn{gop}{G = \langle 0,0 \vert e^{\sum {\b}_{\a}c_{\a}}
II^{\dagger} e^{\sum {\b}_{\a}^{\dagger} c_{\a}^{\dagger}} \vert
0,0\rangle} It is positive definite, which follows from the
representation: $$ G = \langle 0,0 \vert e^{\sum {\b}_{\a}c_{\a}}
\sum {\z}_{\a} ({\b}_{\a}^{\dagger} - c_{\a})({\b}_{\a} -
c_{\a}^{\dagger}) e^{\sum {\b}_{\a}^{\dagger} c_{\a}^{\dagger}}
\vert 0,0\rangle $$ and the fact that ${\b}_{\a} -
c_{\a}^{\dagger}$ has no kernel in ${\CH} \otimes V$. Then define
an element of the algebra ${\CA}_{\t}$ \eqn{prj}{P = I^{\dagger}
e^{\sum {\b}_{\a}^{\dagger} c_{\a}^{\dagger}} \vert 0,0 \rangle
G^{-1} \langle 0,0 \vert e^{\sum {\b}_{\a} c_{\a}} I} which obeys
$P^2 = P$, i.e. it is a projector. Moreover, it is a projection
onto an $N$-dimensional subspace in ${\CH}$, isomorphic to $V$.

\subsubsection{Dual gauge invariance}

The normalization condition \rfn{nrml} is invariant under the
action of the dual gauge group $G_{N} \approx U(N)$ on $B_{\a},
I$. However, the projector $P$ is invariant under the action of
larger group - the complexification
$G_{N}^{\scriptscriptstyle{\IC}} \approx GL_{N}({\IC})$:
\eqn{dualc}{(B_{\a}, I) \mapsto (g^{-1}B_{\a}g, g^{-1}I), \quad
(B_{\a}^{\dagger}, I^{\dagger}) \mapsto
(g^{\dagger}B_{\a}g^{\dagger, -1}, I^{\dagger}g^{\dagger, -1})}
This makes the computations of $P$ possible even when the solution
to the ${\m}_{r} = {\z}_{r}$ part of the ADHM equations is not
known. The moduli space ${\widetilde\CM}_{N,k}$ can be described
both in terms of the hyperkahler reduction as above, or in terms
of the quotient of the space of stable points $Y_{N,k}^{s} \subset
{\m}_{c}^{-1}(0)$ by the action of
$G_{N}^{\scriptscriptstyle{\IC}}$ (see \ctn{donaldson}, \ctn{nakajima} for related discussions). The stable points $(B_1, B_2,
I)$ are the ones where $B_1$ and $B_2$ commute, and generate all
of $V$ by acting on $I$: ${\IC}[ B_1, B_2]\, I = V$, i.e.
precisely those triples which correspond to the codimension $N$
ideals in ${\IC}[z_1, z_2]$.

\subsubsection{Instanton gauge field}
Clearly, $P$ annihilates ${\xi}$, thanks to \rfn{fred}. Let $S$ be
an operator in ${\CH}$ which obeys the following relations:
\eqn{iden}{SS^{\dagger} = 1, \quad S^{\dagger}S = 1- P} The
existence  of $S$ is merely a reflection of the fact that as
Hilbert spaces ${\CH}_{\CI} \approx {\CH}$. So it just amounts to
relabeling the orthonormal bases in ${\CH}_{\CI}$ and ${\CH}$ to
construct $S$.

Now, ${\hat\Delta}$ restricted at the subspace
$S^{\dagger}{\CH}\otimes I \subset {\CH} \otimes V$, is
invertible. We can now solve \rfn{master} as follows:
\eqn{slmaster}{{\xi} = {\Lambda}^{-\half} S^{\dagger}, v = -
{1\over{\hat\Delta}}I {\xi}} where \eqn{mslmbd}{{\Lambda} = 1 +
I^{\dagger}{1\over{\hat\Delta}}I} ${\Lambda}$ is not an element of
${\CA}_{\t}$, but ${\Lambda}^{-1}$ and ${\Lambda}S^{\dagger}$ are.
Finally, the gauge fields can be written as: \eqn{gage}{D_{\a} =
\sqrt{1\over{{\z}_{\a}}} S {\Lambda}^{-\half} c_{\a}
{\Lambda}^{\half} S^{\dagger}, \qquad {\bar D}_{\bar \a} = -
\sqrt{1\over{{\z}_{\a}}} S {\Lambda}^{\half} c_{\a}^{\dagger}
{\Lambda}^{-\half} S^{\dagger}}

\subsubsection{Ideal meaning of $P$}
We can explain the meaning of $P$ in an invariant fashion.
Consider the ideal ${\CI}$ in ${\IC}[z_1, z_2]$, corresponding to
the triple $(B_1, B_2, I)$ as explained above. Any polynomial $f
{\in} {\CI}$ defines a vector $f ( \sqrt{{\z}_1} c_1^{\dagger},
\sqrt{{\z}_2} c_2^{\dagger} ) \vert 0,0 \rangle$ and their
totality span a subspace ${\CH}_{\CI} \subset {\CH}$ of
codimension $N$. The operator $P$ is simply an orthogonal
projection onto the complement to ${\CH}_{\CI}$. The fact ${\CI}$
is an ideal in ${\IC}[z_1, z_2]$ implies that $c_{\a}^{\dagger}
({\CH}_{\CI}) \subset {\CH}_{\CI}$, hence: $$ c_{\a}^{\dagger}
S^{\dagger} {\eta} = S^{\dagger} {\eta}^{\prime} $$ for any
${\eta} \in {\CA}_{\t}$, and also ${\Lambda}^{-\half} S^{\dagger}
= S^{\dagger} {\eta}^{\prime\prime}$ for some ${\eta}^{\prime},
{\eta}^{\prime\prime} \in {\CA}_{\t}$.

Notice that the expressions \rfn{gage} are well-defined. For
example, the ${\bar D}_{\bar \a}$ component contains a dangerous
piece ${\Lambda}^{\half} c_{\a}^{\dagger}\ldots$ in it. However,
in view of the previous remarks it is multiplied by $S^{\dagger}$
from the right and therefore well-defined indeed.

\subsection{Charge one instanton} In this  case: $I = \sqrt{2}$,
one can take $B_{\a} =0$, ${\hat\Delta} = \sum {\z}_{\a} n_{\a}$,
$$ {\Lambda} = {{M + 2}\over{M}}$$ $M = \sum_{\a} {\z}_{\a}
n_{\a}, \, \sum_{\a} {\z}_{\a} = 2$. Let us introduce the notation
$N = n_1 + n_2$. For the pair ${\nb} = (n_1, n_2)$ let $\rho_{\nb}
= {\half}N(N-1) + n_1$. The map ${\nb} \leftrightarrow \rho_{\nb}$
is one-to-one. Let $S^{\dagger} \vert {\rho}_{\nb} \rangle = \vert
{\rho}_{\nb} + 1 \rangle$. Clearly, $SS^{\dagger} = 1, \,
S^{\dagger}S = 1  - \vert 0,0\rangle\langle 0,0 \vert$.

The formulae \rfn{gage} are explicitly non-singular. Let us
demonstrate the anti-self-duality of the gauge field \rfn{gage} in
this case. $$ \sum_{\a} D_{\a}{\bar D}_{\bar\a} = - S
{1\over{{\z}_{\a}}} (n_{\a} + 1) {M\over{M+2}}
{{M+2+{\z}_{\a}}\over{M+{\z}_{\a}}} S^{\dagger} $$ $$ \sum_{\a}
{\bar D}_{\bar\a}D_{\a} = S {1\over{{\z}_{\a}}} n_{\a}
{{M-{\z}_{\a}}\over{M+2-{\z}_{\a}}} {{M+2}\over{M}} S^{\dagger}$$
A simple calculation shows: \eqn{crvaa}{\sum_{\a} [ D_{\a}, {\bar
D}_{\bar\a} ] = - {2\over{{\z}_1{\z}_2}} = - \left(
{1\over{{\z}_1}} + {1\over{{\z}_2}} \right), \qquad [D_{\a},
D_{\b}] = 0,} hence \eqn{asdcrva}{\sum_{\a} F_{\a{\bar\a}} = 0} as
$$ i\sum_{\a} {\t}_{\a\bar\a} = {1\over{{\z}_1}} +
{1\over{{\z}_2}} $$ This is a generalization of the charge one
instanton constructed in \ctn{neksch}, written in the explicitly
non-singular gauge.

\subsubsection{Remark on gauges}

The gauge which was chosen in the examples considered in
\ctn{neksch} and subsequently adopted in \ctn{fki}, \ctn{gms} had
${\xi} = {\xi}^{\dagger}$. It was shown in \ctn{fki} that this
gauge does not actually lead to the canonically normalized
identificator ${\Psi}$: one had ${\Psi}^{\dagger}{\Psi} = 1 - P$.
Our paper showed that this gauge is in some sense an analogue of
't Hooft singular gauge for commutative instantons: it leads to
singular formulae, if the gauge field is considered to be
well-defined globally over ${\CA}_{\t}$. However, as we showed
above, there are gauges in which the gauge field is globally
well-defined, non-singular, and anti-self-dual. They simply have
${\xi} \neq {\xi}^{\dagger}$.

\section*{Acknowledgements}

I would like to thank H.~Braden, D.~Gross and S.~Shatashvili for
useful discussions. I would also like to thank E.~Witten for 
stressing the point that 
\nct instantons must exist without any space topology change.
 
The research was partially supported by NSF
under the grant PHY94-07194, by Robert H.~Dicke fellowship from
Princeton University, partly by RFFI under grant 00-02-16530,
partly by the grant 00-15-96557 for scientific schools. I thank
the ITP at University of California, Santa Barbara, and CIT-USC
center for hospitality while this paper was written.

\end{document}